\documentclass[twocolumn,showpacs,preprintnumbers,amsmath,amssymb,prl,superscriptaddress]{revtex4}

\usepackage{graphicx}
\usepackage{dcolumn}
\usepackage{bm}
\usepackage{color} 

\begin{document}

\title{Electronic structure of semiconducting CeFe$_4$P$_{12}$: Strong hybridization and relevance of single-impurity Anderson model}

\author{M.~Matsunami}
 \altaffiliation[Electronic address: ]{matunami@spring8.or.jp}
\author{K.~Horiba}
\author{M.~Taguchi}
\author{K.~Yamamoto}
\author{A.~Chainani}
\author{Y.~Takata}
\affiliation{Soft X-Ray Spectroscopy Laboratory, RIKEN SPring-8 Center, Sayo-cho, Sayo-gun, Hyogo 679-5148, Japan}
\author{Y.~Senba}
\author{H.~Ohashi}
\affiliation{JASRI/SPring-8, Sayo-cho, Sayo-gun, Hyogo 679-5148, Japan}
\author{M.~Yabashi}
\affiliation{JASRI/SPring-8, Sayo-cho, Sayo-gun, Hyogo 679-5148, Japan}
\affiliation{Coherent X-Ray Optics Laboratory, RIKEN SPring-8 Center, Sayo-cho, Sayo-gun, Hyogo 679-5148, Japan}
\author{K.~Tamasaku}
\author{Y.~Nishino}
\author{D.~Miwa}
\affiliation{Coherent X-Ray Optics Laboratory, RIKEN SPring-8 Center, Sayo-cho, Sayo-gun, Hyogo 679-5148, Japan}
\author{T.~Ishikawa}
\affiliation{JASRI/SPring-8, Sayo-cho, Sayo-gun, Hyogo 679-5148, Japan}
\affiliation{Coherent X-Ray Optics Laboratory, RIKEN SPring-8 Center, Sayo-cho, Sayo-gun, Hyogo 679-5148, Japan}
\author{E.~Ikenaga}
\author{K.~Kobayashi}
\affiliation{JASRI/SPring-8, Sayo-cho, Sayo-gun, Hyogo 679-5148, Japan}
\author{H.~Sugawara}
\affiliation{Faculty of Integrated Arts and Sciences, The University of Tokushima, Tokushima 770-8502, Japan}
\author{H.~Sato}
\affiliation{Department of Physics, Tokyo Metropolitan University, Hachioji, Tokyo 192-0397, Japan}
\author{H.~Harima}
\affiliation{Department of Physics, Kobe University, Kobe 657-8501, Japan}
\author{S.~Shin}
\affiliation{Soft X-Ray Spectroscopy Laboratory, RIKEN SPring-8 Center, Sayo-cho, Sayo-gun, Hyogo 679-5148, Japan}
\affiliation{Institute for Solid State Physics, The University of Tokyo, Kashiwa, Chiba 277-8581, Japan}

\date{\today}

\begin{abstract} 
Semiconducting skutterudite CeFe$_4$P$_{12}$ is investigated by synchrotron x-ray photoemission spectroscopy (PES) and x-ray absorption spectroscopy (XAS). 
Ce 3$d$ core-level PES and 3$d$-4$f$ XAS, in combination with single impurity Anderson model (SIAM) calculations, confirm features due to $f^0$, $f^1$ and $f^2$ configurations. 
The Ce 4$f$ density of states (DOS) indicates absence of a Kondo resonance at Fermi level, but can still be explained by SIAM with a small gap in non-$f$ DOS.  While Ce 4$f$ partial DOS from band structure calculations are also consistent with the main Ce 4$f$ DOS, the importance of SIAM for core and valence spectra indicates Kondo semiconducting mixed valence for CeFe$_4$P$_{12}$, derived from strong hybridization between non-$f$ conduction and Ce 4$f$ DOS. 

\end{abstract}



\maketitle
Filled skutterudite compounds with a general formula $RT_4X_{12}$ ($R$ = rare earths; $T$ = Fe, Ru, Os; $X$ = P, As, Sb) exhibit a wide variety of strongly correlated electron phenomena, e.g., exotic superconductivity: PrOs$_4$Sb$_{12}$ \cite{PrOs4Sb12}; heavy fermion behavior: PrFe$_4$P$_{12} $\cite{PrFe4P12}; metal-insulator transition: PrRu$_4$P$_{12}$\cite{PrRu4P12} and SmRu$_4$P$_{12}$ \cite{SmRu4P12}; unusual non-metallicity: CeFe$_4$P$_{12}$ and UFe$_4$P$_{12}$ \cite{CeFe4P12&UFe4P12}, etc. 
The filled skutterudites are also known to be important for thermoelectric applications \cite{Thermoelectric}. 
The properties depending on rare earth ion $R$ in terms of its ionic size, electronic states, and its position in the unique bcc crystal structure (space group $Im\bar3$, with $R$ ions surrounded by twelve $X$ and eight $T$ ions), suggest a tunable hybridization between the rare-earth $f$-electrons and non-$f$ conduction ($c$) band states as primarily responsible for this variety of phenomena \cite{Sales,Aoki}.

Among the $R$Fe$_4$P$_{12}$ series with $R$ belonging to the 4$f$ series (R = La, Ce, Pr, Nd), only the Ce-compound is semiconducting; 
the LaFe$_4$P$_{12}$ is a superconductor ($T_{\rm c}$ = 4.1~K) while the  PrFe$_4$P$_{12}$ and NdFe$_4$P$_{12}$ show metallic behavior \cite{CeFe4P12&UFe4P12,RFe4P12_base,CeFe4P12_base}. 
Even for single-crystals of CeFe$_4$P$_{12}$, the fit to an activated behavior is valid only over a limited temperature range and gives a gap value of $\sim$0.13 eV, consistent with a 0.15~eV gap-like structure in optical spectroscopy \cite{CeFe4P12_optical}. 
The temperature independent magnetic susceptibility is also anomalous, with a value approximately half of that of LaFe$_4$P$_{12}$. 
This fact indicates a reduced magnetic moment for Ce ions, since the metallic PrFe$_4$P$_{12}$ and NdFe$_4$P$_{12}$ exhibit typical trivalent free-ion magnetic moments. 
The above described behavior of CeFe$_4$P$_{12}$ is thus, similar to the hybridization gap semiconductors which are also referred to as Kondo-semiconductors \cite{KondoSemi_Theory1}. 
In particular, among Ce-filled skutterudites, the Ce$T$$_4$P$_{12}$ series show semiconducting properties and have much smaller lattice constants than the value expected from trivalent lanthanide contraction \cite{CeT4P12_LC}. 
This itself suggests that Ce 4$f$ states have strong hybridization with $c$ electron states, and hence the energy gap in Ce$T_4$P$_{12}$ may be a $c$-$f$ hybridization gap. 
While a few studies on Ce$T_4$P$_{12}$ \cite{CeT4P12_base,CeRu4P12_XANES} support this picture, there is no direct experimental evidence for the strong $c$-$f$ hybridization in these compounds.

In this work, we use core-level photoemission spectroscopy (CL-PES), 3$d$-4$f$ resonant-PES (R-PES) and x-ray absorption spectroscopy (XAS) to investigate the electronic structure of CeFe$_4$P$_{12}$. 
PES and XAS, in combination with single-impurity Anderson model (SIAM) calculations, have played a major role in studying the electronic structure of strongly correlated metallic $f$ electron systems exhibiting the Kondo effect, mixed valence and heavy fermion behavior \cite{GS,Allen}. 
These techniques provide a quantification of electronic structure parameters as well as the $f$-electron count, $n_{f}$, which is a direct measure of hybridization induced mixed valency. 
Recent work has also shown that soft x-ray (SX) 3$d$-4$f$ R-PES \cite{Bulk_Sekiyama} and hard x-ray (HX)-PES \cite{HX-PES_Inst1,HX-PES_Inst2,Taguchi} are important for studying bulk-sensitive electronic structure of solids, while 4$d$-4$f$ R-PES probes surface sensitive electronic structure due to lower escape depths.

High-resolution low energy studies of Kondo semiconductors have identified a clear pseudogap behavior at Fermi level ($E_{\rm F}$) in YbB$_{12}$ \cite{Susaki}, as well as in CeRhAs and CeRhSb \cite{Kumigashira}. 
Recent studies of 3$d$-4$f$ R-PES \cite{Sekiyama} concluded a bulk pseudogap in CeRhAs, while valence band HX-PES \cite{Shimada} with $h\nu$ $\sim$ 6~keV of CeRhAs were analysed in terms of non-$f$ DOS, since $f$-electron photoionization cross-section is strongly reduced at $h\nu $$\sim$ 6~keV. 
Thus, methods of electron spectroscopy combined with SIAM are well-established for typical Kondo metals \cite{GS,Allen}, to date. 
However, high-energy CL-PES in combination with 3$d$-4$f$ XAS and R-PES, and their consistency with SIAM calculations for a hybridization gap semiconductor have not been established.


CeFe$_4$P$_{12}$ single crystals used in this work were grown by the tin-flux method \cite{CeFe4P12_base}. 
 SX-PES (both CL- and R-PES), x-ray absorption spectroscopy (XAS) and HX-CL-PES were performed at SPring-8. 
For all experiments, a clean surface was obtained by fracturing $in~situ$ at low temperatures. 
The vacuum during measurements was around $2\times10^{-8}$~Pa. 
SX-PES and XAS experiments were performed at undulator beamline BL17SU. 
PES spectra were measured using a hemispherical electron analyzer, SCIENTA SES-2002. 
The overall energy resolution was set to $\sim$ 200~meV for the spectra of CL-PES and R-PES. 
XAS spectra were recorded using the total electron yield method. 
HX-PES ($h\nu$ = 5948~eV) experiments were performed at undulator beamline BL29XU using a hemispherical electron analyzer, SCIENTA SES-2002 \cite{HX-PES_Inst2} at
a total energy resolution of $\sim$ 300~meV.

\begin{figure}[t]
\begin{center}
\includegraphics[width=0.38\textwidth]{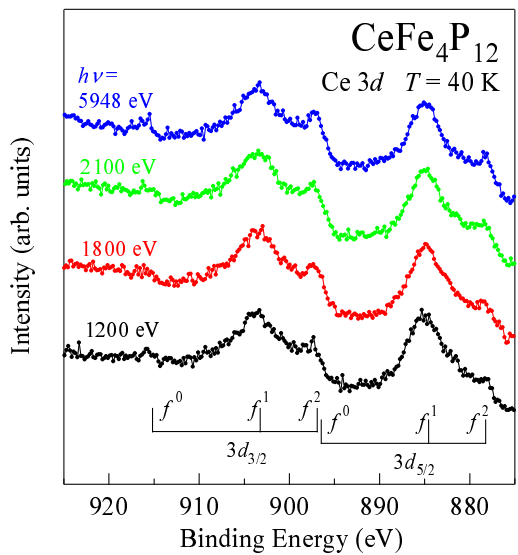}
\caption{
(Color online) Ce 3$d$ core-level photoemission spectra for CeFe$_4$P$_{12}$ measured at several excitation energies. 
} 
\end{center}
\end{figure}


Figure 1 shows Ce 3$d$ CL-PES spectra of CeFe$_4$P$_{12}$ measured at excitation energies from 1200 to 5948 eV. 
The excitation energy dependent spectra are normalized for the intensity of the 3$d_{5/2}f^1$ peak. 
The spectra show common features, with a 3-peak structure labelled as $f^0$, $f^1$ and $f^2$ states, for the spin-orbit split Ce 3$d_{5/2}$ and 3$d_{3/2}$ energy ranges separated by about 18~eV. 
The 3$d_{5/2} f^0$ and 3$d_{3/2} f^2$ final states overlap. 
The 3-peak structure is typical of Ce-based Kondo or mixed valent metals and is thus surprising, as CeFe$_4$P$_{12}$ is a semiconductor. 
The presence of pronounced $f^0$ and $f^2$ final states suggests that screening effects in CL-PES of CeFe$_4$P$_{12}$ are qualitatively similar to Kondo metals. 
In particular, since the $f^2$ final state is caused by a screening of the core hole by electrons from the valence band to 4$f$ states, 
the $f^2$ peak intensity is considered as an indication of $c$-$f$ hybridization strength. 
This evidence for valence fluctuation derived from strong $c$-$f$ hybridization, is quantified in the following, using SIAM calculations. 
With increasing excitation energy, i.e., increasing bulk-sensitivity, the $f^2$ and $f^0$ peak intensities show a slight increase from $h\nu$ = 1200~eV to 2100~eV (as seen in 3$d_{5/2}$-$f^2$ peak), but negligible change for $h\nu$ = 2100~eV and 5948~eV. 
This result indicates that the spectra are dominated by bulk-derived states at and above $h\nu$ = 2100~eV. 
A similar behavior has been reported in a series of Ce-based Kondo metals \cite{CeM2_XPS}.

\begin{figure}[t]
\begin{center}
\includegraphics[width=0.38\textwidth]{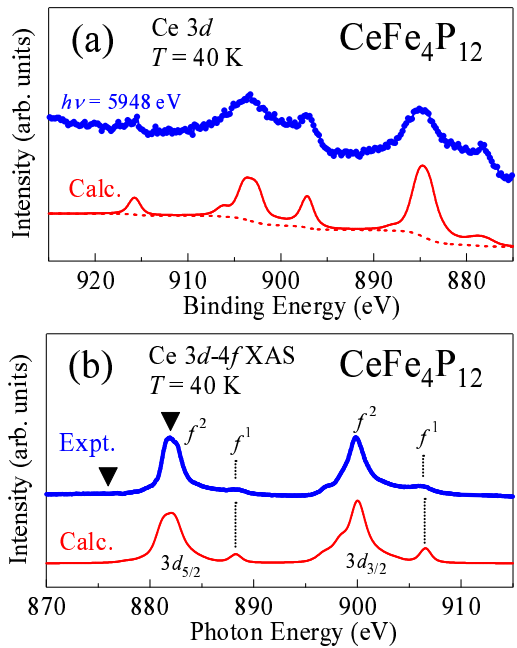}
\caption{
(Color online) Comparison of (a) Ce 3$d$ core-level photoemission spectrum and (b) Ce 3$d$-4$f$ x-ray absorption spectrum with corresponding spectra calculated by the SIAM. 
Solid triangles mark energies at which off and on-resonant photoemission spectra in Fig. 3 were measured. 
} 
\end{center}
\end{figure}


In order to obtain the electronic structure parameters of CeFe$_4$P$_{12}$, we use the modified SIAM with full multiplet effects to calculate the Ce 3$d$ CL-PES spectrum. 
We use the basis set consisting $4f^0$, $4f^1\underline{v}$, and $4f^2\underline{v}^2$ configurations to describe the ground state, where $\underline{v}$ is a hole in the $c$ band below $E_{\rm F}$. 
The hybridization strength $V(\varepsilon)$ between the $4f$ and $c$ band depends on the $c$ band energy $\varepsilon$. 
The $c$ band states are logarithmically discretized \cite{nak02}, with a semi-elliptical form of $V(\varepsilon)^2$ and a small gap (0.2~eV) at $E_{\rm F}$ (Fig.~4), because CeFe$_4$P$_{12}$ is a semiconductor. 
Details of the model calculations are standard and are described in earlier work \cite{GS,nak02,Rc&Rv}. 
Figure~2 (a) shows a comparison of the Ce 3$d$ CL-PES spectrum measured at $h\nu$ = 5948~eV with the calculated spectrum obtained using the SIAM. 
The calculated spectra match the experimental spectra fairly well and the estimated parameters are: 
on-site Coulomb repulsion $U_{ff}$ = 7~eV, 
core-hole potential $U_{fc}$ = 11.78~eV, 
charge-transfer energy $\Delta$ = $-$0.9~eV, 
hybridization strength $V$ = 0.44~eV, 
bandwidth $W$ = 2.0~eV and $n_f$ = 0.86. 
These values are similar to those of $\alpha$-Ce investigated by bulk sensitive spectroscopy \cite{Ce_RIXS}. 
The results clearly indicate mixed valency due to the strong $c$-$f$ hybridization in CeFe$_4$P$_{12}$. 
Using the same parameter set with a small change only in $V$ (= 0.4~eV), 
we have also calculated the Ce 3$d$-4$f$ XAS spectrum and compared with experiments, as shown in Fig.~2 (b). 
For the XAS spectrum, weak satellite structures at 888~eV and 907~eV are clearly observed, with a slightly lower intensity in the experiment compared to the calculation. 
Nonetheless, these weak satellites are due to the 3$d^9$4$f^1$ final state originating from the 4$f^0$ configuration, and indicate the importance of the $f^0$ contribution in the ground state of CeFe$_4$P$_{12}$. 
A small reduction in $V$ is known for XAS of Ce-based Kondo metals from early work \cite{GS}.


\begin{figure}[t]
\begin{center}
\includegraphics[width=0.39\textwidth]{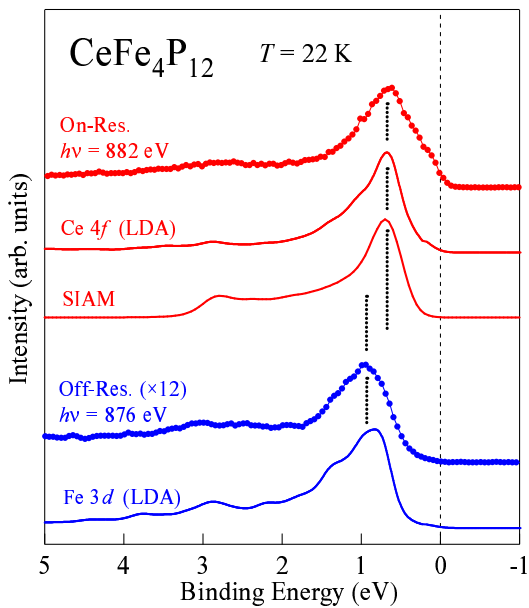}
\caption{
(Color online) Ce 3$d$-4$f$ resonant photoemission spectra for CeFe$_4$P$_{12}$ in comparison with the LDA band structure calculation (Ce 4$f$ pDOS) and SIAM. 
Off-resonant spectrum is compared with LDA Fe 3$d$ pDOS. 
} 
\end{center}
\end{figure}

Figure 3 shows the off- and on-R valence band PES spectra of CeFe$_4$P$_{12}$, measured at $h\nu$ = 876~eV and 882~eV, respectively. 
The photon energies used are marked in the XAS spectrum shown in Fig.~2 (b). 
The spectra are normalized to the incident photon flux. 
The large increase (about twelvefold) in spectral intensity of on-R PES spectrum compared to off-R PES spectrum represents the Ce 4$f$ partial density of states (pDOS). In typical Ce-based Kondo metals, the Ce 4$f$ pDOS obtained from the on-R PES spectrum is well-described by SIAM calculations \cite{GS,Allen,Bulk_Sekiyama}. 
In this model, the Ce 4$f$ pDOS in the occupied DOS is composed of $f^0$ final state at high binding energy ($\sim$3 eV), and $f^1$ final state corresponding to the so-called Kondo resonance peak just at and above $E_{\rm F}$ \cite{GS,Allen,Bulk_Sekiyama}. The intensity of $f^1$ peak at and near $E_{\rm F}$ is enhanced by electron-hole pair excitations \cite{nak02}. 
In contrast, the experimentally obtained Ce 4$f$ pDOS exhibits a main peak significantly below $E_{\rm F}$, positioned at 0.7 eV binding energy. 
Thus CeFe$_4$P$_{12}$ does not behave like a regular Kondo metal as seen by the absence of the Kondo resonance at $E_{\rm F}$, although Ce 3$d$ CL-PES can be described by the SIAM. 
We believe this is in accord with its semiconducting property. 
Figure~4 schematically shows $f$-level and $c$ states for a regular Kondo metal and its modification due to a semiconducting gap. 
The presence of a small gap still allows screening of a core-hole from occupied valence band states giving characteristic features observed in CL-PES and XAS, but does not give the Kondo resonance peak. 
Consequently, the Ce 4$f$ features result in qualitatively different behavior, unlike a conventional SIAM calculation for a Kondo metal. 
Therefore, we introduce a modified SIAM model, which has a narrow energy gap at $E_{\rm F}$ (Fig.4). 
The parameters used for the calculation of on-R PES spectrum are the same as that of XAS. 
The calculated spectrum shows a reasonable agreement with experimental on-R PES spectrum, except for a weak structure just below $E_{\rm F}$. 
Note that while the Ce 3$d$ CL-PES is like $\alpha$-Ce, the Ce 4$f$ states are very different from $\alpha$-Ce. 
The weak structure in on-R PES spectrum also has 4$f$ character as its spectral intensity is suppressed in the off-R PES spectrum (Fig. 3).

\begin{figure}[t]
\begin{center}
\includegraphics[width=0.45\textwidth]{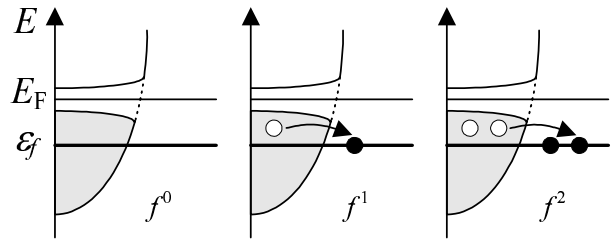}
\caption{
Schematic for the ground state configurations in the SIAM with an energy gap at Fermi level. 
Dashed lines : normal metal DOS (no energy gap). 
} 
\end{center}
\end{figure}

We have also compared the on- and off-R PES spectra with band structure calculations using the local density approximation (LDA) \cite{Band_Calc_Harima}, as shown in Fig.~3. 
The valence and conduction bands around $E_{\rm F}$ in CeFe$_4$P$_{12}$ are composed of Ce 4$f$, Fe 3$d$ and P 3$p$ states. 
Taking into account the photoionization cross sections \cite{CrossSection}, the off-R PES spectrum is expected to be dominated by Fe 3$d$ pDOS, while the on-R data is dominated by Ce 4$f$ pDOS. 
Except for a weak feature around $E_{\rm F}$ in the on-R PES spectrum, the Ce 4$f$ spectral shape is also well reproduced by the LDA calculation. 
It is surprising that the Ce 4$f$ electronic states are similarly reproduced by both SIAM and LDA calculations, since the calculations are generally incompatible with each other \cite{CeMx}. 
The off-R PES spectrum is nicely reproduced by the calculated Fe 3$d$ pDOS. 
The main peaks of the Fe 3$d$ and Ce 4$f$ appear at $\sim$ 1~eV and $\sim$ 0.7~eV, respectively, and this difference in the peak positions is quantitatively agree with the Fe 3$d$ and Ce 4$f$ calculated pDOS. 
This result suggests that the majority of states near $E_{\rm F}$ in CeFe$_4$P$_{12}$ has Ce 4$f$ character. 
According to $c$-$f$ hybridization model \cite{KondoSemi_Theory1} for Kondo semiconductors, the hybridization between the narrow $f$ band positioned just below $E_{\rm F}$ and a broad $c$ band produces an energy gap at $E_{\rm F}$. 
The CL-PES and XAS results give convincing evidence for strong hybridization in CeFe$_4$P$_{12}$, with a $n_{f}$ = 0.86 typical of mixed valent $\alpha$-Ce type systems. 
A cluster model analysis of PES spectra for PrFe$_4$P$_{12}$ \cite{PES_PrFe4P12}, concluded a heavy fermion metallic behavior due to strong hybridization with an $n_f$ count of 2.07, corresponding to divalent admixture \cite{Kucherenko}. 
Therefore, CeFe$_4$P$_{12}$ is ever more mixed valent and has stronger hybridization than PrFe$_4$P$_{12}$. 
The experimental data lends strong credence to the gap in CeFe$_4$P$_{12}$ being a $c$-$f$ hybridization gap. 
Thus, while SIAM and LDA calculations describe the Ce 4$f$ pDOS equally well, the features derived from $f^0$, $f^1$ and $f^2$ configurations in CL-PES and XAS indicate a Kondo semiconducting ground state for CeFe$_4$P$_{12}$.


In summary, we have performed synchrotron PES and XAS on the filled skutterudite CeFe$_4$P$_{12}$. 
The Ce 3$d$ CL-PES and XAS, in combination with SIAM calculations, indicate mixed valency in CeFe$_4$P$_{12}$ due to strong $c$-$f$ hybridization. 
Ce 3$d$-4$f$ R-PES reveals the absence of a Kondo resonance, consistent with a semiconducting ground state. 
While band structure and SIAM calculations show qualitative agreement with the main peak of Ce 4$f$ pDOS, features derived from $f^0$, $f^1$ and $f^2$ configurations in CL-PES and XAS clearly indicate a Kondo semiconducting ground state of CeFe$_4$P$_{12}$. 


The authors thank Drs. T.~Takeuchi and R.~Eguchi for technical support. This work was supported by a Grant-in-Aid for Scientific Research Priority Area "Skutterudite" (Nos. 15072203 and 15072206) of the Ministry of Education, Culture, Sports, Science and Technology, Japan.



\begin{thebibliography}{99}

\bibitem{PrOs4Sb12} 
E.~D.~Bauer $et~al$., Phys. Rev. B {\bf 65}, 100506(R) (2002). 

\bibitem{PrFe4P12} 
H.~Sugawara $et~al$., 
Phys. Rev. B {\bf 66}, 134411 (2002). 

\bibitem{PrRu4P12} 
C.~Sekine $et~al$., 
Phys. Rev. Lett. {\bf 79}, 3218 (1997). 

\bibitem{SmRu4P12} 
C.~Sekine $et~al$., 
Science and Technology of High Pressure, Universities Press, Hyderabad, 
India, 826 (2000). 

\bibitem{CeFe4P12&UFe4P12} 
G.~P.~Meisner $et~al$., 
J. Apple. Phys. {\bf 57}, 3073 (1985). 

\bibitem{Thermoelectric}
B.~C.~Sales, D.~Mandrus and R.~K.~Williams, 
Science {\bf 272}, 1325 (1996). 

\bibitem{Sales}
B.~C.~Sales, in Handbook of Physics and Chemistry of the Rare Earths, Vol.33 Ch. 211, eds. K.A. Gschneidner, J.~-C.~Bunzli and V.~K.~Pecharsky (2003). 

\bibitem{Aoki}
Y.~Aoki $et~al$., 
J. Phys. Soc. Jpn. {\bf 74}, 209 (2005). 

\bibitem{RFe4P12_base}
M.~S.~Torikachvili $et~al$., 
Phys. Rev. B {\bf 36}, 8660 (1987). 

\bibitem{CeFe4P12_base} 
H.~Sato $et~al$., 
Phys. Rev. B {\bf 62}, 15125 (2000). 

\bibitem{CeFe4P12_optical} 
S.~V.~Dordevic $et~al$., 
Phys. Rev. B {\bf 60}, 11321 (1999). 

\bibitem{KondoSemi_Theory1} 
G.~Aeppli and Z.~Fisk, 
Comments Condens. Matter Phys. {\bf 16}, 155 (1992). 

\bibitem{CeT4P12_LC} 
F.~Grandjean $et~al$., 
J. Phys. Chem. Solids {\bf 45}, 877 (1984). 

\bibitem{CeT4P12_base} 
I.~Shirotani $et~al$., 
J. Solid State Chem. {\bf 142}, 146 (1999). 

\bibitem{CeRu4P12_XANES} 
C.~H.~Lee $et~al$., 
Phys. Rev. B {\bf 60}, 13253 (1999). 

\bibitem{GS}
O.~Gunnarsson and K.~Sch{\"o}nhammer, 
Phys. Rev. B {\bf 31}, 4815 (1985); 
Phys. Rev. B {\bf 28}, 4315 (1985). 

\bibitem{Allen}
J. W. Allen $et~al$., 
Adv. Phys. {\bf 35}, 275 (1986). 

\bibitem{Bulk_Sekiyama} 
A.~Sekiyama $et~al$., 
Nature {\bf 403}, 396 (2000). 

\bibitem{HX-PES_Inst1} 
K.~Kobayashi $et~al$., 
Appl. Phys. Lett. {\bf 83}, 1005 (2003). 

\bibitem{HX-PES_Inst2} 
Y.~Takata $et~al$., 
Appl. Phys. Lett. {\bf 84}, 4310 (2004). 

\bibitem{Taguchi}
M.~Taguchi $et~al$., 
Phys. Rev. Lett. {\bf 95}, 177002 (2005). 

\bibitem{Susaki} 
T.~Susaki $et~al$., 
Phys. Rev. Lett. {\bf 82}, 992 (1999).

\bibitem{Kumigashira} 
H.~Kumigashira $et~al$., 
Phys. Rev. Lett. {\bf 87}, 067206 (2001).

\bibitem{Sekiyama} 
A.~Sekiyama $et~al$., 
Physica B {\bf 359-361}, 115 (2005).

\bibitem{Shimada} 
K.~Shimada $et~al$., 
Nucl. Inst. Methods Phys. Res. {\bf 547}, 169 (2005).

\bibitem{CeM2_XPS} 
L.~Braicovich $et~al$., 
Phys. Rev. B {\bf 56}, 15047 (1997). 

\bibitem{nak02}
M.~Nakazawa and A.~Kotani, 
J. Phys. Soc. Jpn. {\bf 71}, 2804 (2002). 

\bibitem{Rc&Rv} 
In the calculations, the $c$-$f$ hybridization is reduced by a factor $R_{\rm C}$ (=0.7) in the presence of a core hole and enhanced by a factor $1/R_{\rm V}$ ($R_{\rm V} = 0.7$) in the presence of an extra 4$f$ electron. 
The framework is discussed in detail in 
O.~Gunnarsson and O.~Jepsen, 
Phys. Rev. B {\bf 38}, R3568 (1988). 

\bibitem{Ce_RIXS} 
C.~Dallera $et~al$., 
Phys. Rev. B {\bf 70}, 085112 (2004). 

\bibitem{CrossSection} 
J.~J.~Yeh and I.~Lindau, At. Data Nucl. Data Tables {\bf 32}, 1 (1985). 

\bibitem{Band_Calc_Harima} 
H.~Harima, unpublished. 

\bibitem{CeMx}
R.-J.~Jung $et~al$., 
Phys. Rev. Lett. {\bf 91}, 157601 (2003). 

\bibitem{PES_PrFe4P12}
A.~Yamasaki $et~al$., Phys. Rev. B {\bf 70}, 113103 (2004); 
A.~Yamasaki $et~al$., J. Phys. Soc. Jpn. {\bf 74}, 2045 (2005). 

\bibitem{Kucherenko}
Y. Kucherenko $et~al$., 
Phys. Rev. B {\bf 66}, 165438 (2002). 



\end{thebibliography}

\end{document}